\documentclass[12pt]{iopart}
\usepackage[utf8]{inputenc}
\usepackage{graphicx}
\usepackage{braket}
\usepackage{mathtools}
\usepackage[colorlinks=true,citecolor=blue,linkcolor=blue,urlcolor=blue]{hyperref}
\usepackage{cite}

\begin{document}

\title{Variational Quantum Eigensolver for SU(\textit{N}) Fermions}

\author{Mirko Consiglio}
\ead{mirko.consiglio.16@um.edu.mt}
\address{Department of Physics, University of Malta, Msida MSD 2080, Malta}

\author{Wayne J. Chetcuti}
\address{Dipartimento di Fisica e Astronomia, Via S. Sofia 64, 95127 Catania, Italy}
\address{INFN-Sezione di Catania, Via S. Sofia 64, 95127 Catania, Italy}
\address{Quantum Research Centre, Technology Innovation Institute, Abu Dhabi, UAE}

\author{Carlos Bravo-Prieto}
\address{Quantum Research Centre, Technology Innovation Institute, Abu Dhabi, UAE}
\address{Departament de Física Quàntica i Astrofísica and Institut de Ciències del Cosmos (ICCUB), Universitat de Barcelona, Martí i Franquès 1, 08028 Barcelona, Spain.}

\author{Sergi Ramos-Calderer}
\address{Quantum Research Centre, Technology Innovation Institute, Abu Dhabi, UAE}
\address{Departament de Física Quàntica i Astrofísica and Institut de Ciències del Cosmos (ICCUB), Universitat de Barcelona, Martí i Franquès 1, 08028 Barcelona, Spain.}

\author{Anna Minguzzi}
\address{Université Grenoble Alpes, CNRS, LPMMC, 38000 Grenoble, France}

\author{José I. Latorre}
\address{Quantum Research Centre, Technology Innovation Institute, Abu Dhabi, UAE}
\address{Departament de Física Quàntica i Astrofísica and Institut de Ciències del Cosmos (ICCUB), Universitat de Barcelona, Martí i Franquès 1, 08028 Barcelona, Spain.}
\address{Centre for Quantum Technologies, National University of Singapore, 3 Science Drive 2, Singapore 117543, Singapore}

\author{Luigi Amico}
\address{Quantum Research Centre, Technology Innovation Institute, Abu Dhabi, UAE}
\address{Centre for Quantum Technologies, National University of Singapore, 3 Science Drive 2, Singapore 117543, Singapore}
\address{CNR-MATIS-IMM \& INFN-Sezione di Catania, Via S. Sofia 64, 95127 Catania, Italy}
\address{LANEF `Chaire d’excellence', Université Grenoble Alpes, CNRS, 38000 Grenoble, France}

\author{Tony J. G. Apollaro}
\ead{tony.apollaro@um.edu.mt}
\address{Department of Physics, University of Malta, Msida MSD 2080, Malta}

\begin{abstract}
Variational quantum algorithms aim at harnessing the power of noisy intermediate-scale quantum computers, by using a classical optimizer to train a parameterized quantum circuit to solve tractable quantum problems. The variational quantum eigensolver is one of the aforementioned algorithms designed to determine the ground-state of many-body Hamiltonians. Here, we apply the variational quantum eigensolver to study the ground-state properties of $N$-component fermions. With such knowledge, we study the persistent current of interacting SU($N$) fermions, which is employed to reliably map out the different quantum phases of the system. Our approach lays out the basis for a current-based quantum simulator of many-body systems that can be implemented on noisy intermediate-scale quantum computers.
\end{abstract}

\section{Introduction}

The current information technology era is being reshaped by emerging quantum technologies in fields ranging from cryptography to computation. In this context, noisy intermediate-scale quantum (NISQ) computers may already provide useful applications before fault-tolerant quantum computation is realized \cite{Preskill2018a, Bharti2021}.

Interacting many-body ground-states and quantum phases of matter define important domains in which NISQ devices could play a pivotal role. The necessary steps in order to address a many-body problem via a quantum computer consist of representing the system under scrutiny in terms of qubits, and executing an efficient algorithm to compute the quantity of interest. The system's specific qubit representation has to be devised carefully to optimize the complexity of the resulting quantum circuit. Similarly, a great variety of algorithms have been proposed, and variational quantum algorithms are characteristic examples in the NISQ era exploiting the quantum advantage provided by few-body, noisy quantum computers working alongside classical computers \cite{Cerezo2020}.

\begin{figure}[!ht]
	\centering
	\includegraphics[width=\textwidth]{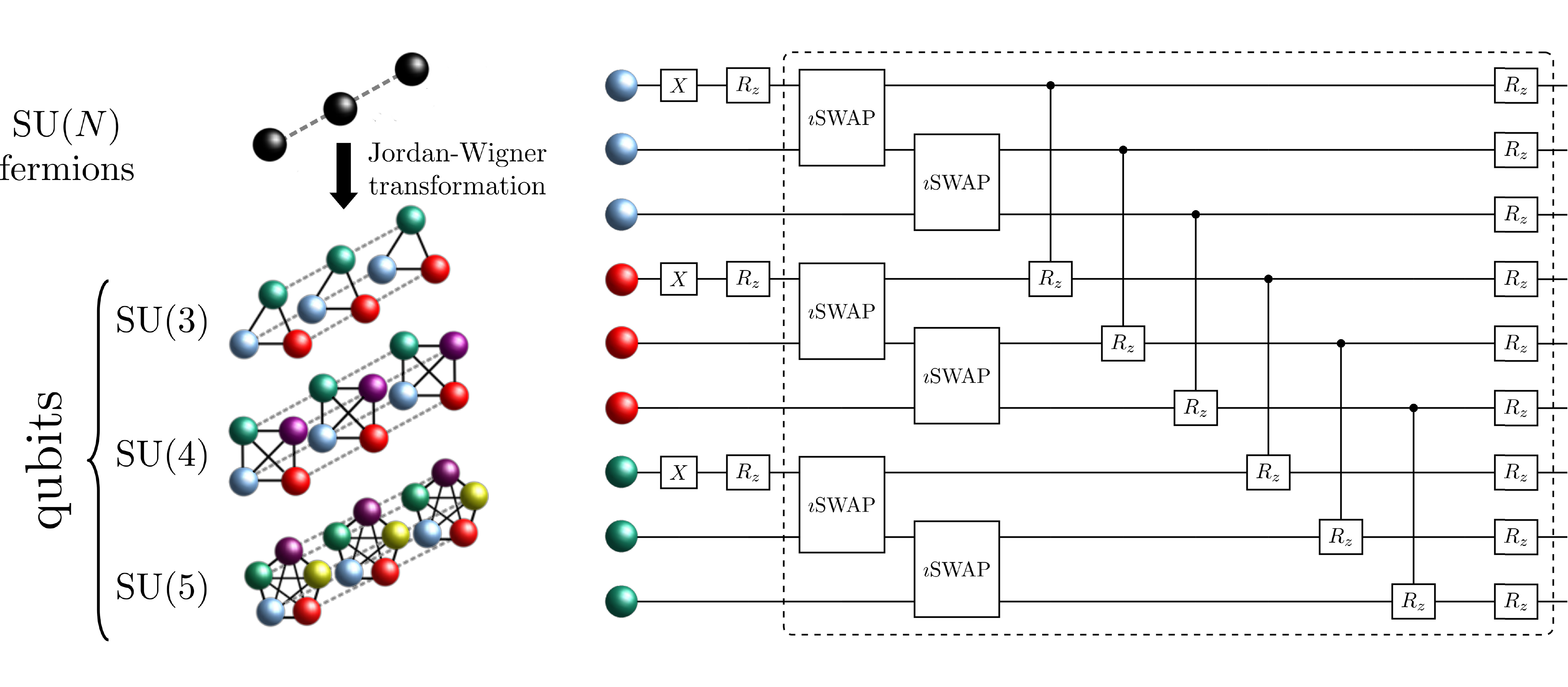}
	\caption{Parameterized quantum circuit employed for the VQE. (Left) Mapping an 3-site linear SU($N$) Hubbard Hamiltonian (black spheres) with only nearest-neighbor hopping and on-site interaction to an $N$-sided qubit prism of length $L$. The dashed gray lines represent the hopping term $t$ between same-color qubits, and continuous black lines the on-site interaction $U$. (Right) Parameterized quantum circuit for the 3-site SU(3) Hubbard Hamiltonian in the $N_p=3$ (number of particles) sector starting from a state with all fermions on the same site (which are created in the circuit via initial $X$ gates). The variational parameters of the parameterized quantum circuit are contained within the $\imath$SWAP, controlled-$R_z$ and $R_z$ gates. The dashed box corresponds to one variational circuit layer. Note that the last sequence of entangling gates can be mostly applied in parallel.}
	\label{fig:circuit}
\end{figure}

The Hubbard model, originally introduced to study the dynamics of electrons in solids \cite{Hubbard1963}, is a paradigmatic example in addressing the physical properties of strongly interacting quantum many-body systems, ranging from superconductivity to quantum magnetism \cite{essler_2005}. The Hubbard model describes itinerant electrons sensing a local interaction. With cold atom quantum technology, the Hubbard model can be studied with unprecedented control, and flexibility of the system's physical conditions \cite{Lewenstein2007, Mazurenko2017, Tarruell2018,esslinger2010fermi}. Despite the simple logic implied in the Hubbard Hamiltonian dynamics, finding its ground-state poses a challenging problem in many-body physics, whose solution has been attempted numerous times by exploiting the most advanced available methods \cite{LeBlanc2015}. With no exception, different quantum algorithms have been devised to address the problem \cite{hensgens2017quantum,Cade2020, Cai2020,Abrams1997, Dallaire-Demers2016,Bauer2016, Reiner_2019, uvarov2020variational}, mainly focusing on determining the ground-state via the application of the aforementioned variational quantum algorithms.

Here, we apply the variational quantum eigensolver (VQE) \cite{peruzzo2014variational} to SU($N$) Hubbard type models, describing strongly interacting fermions with $N$ spin-components. Interacting SU($N$) fermions play an important role in a variety of different contexts ranging from high energy \cite{Cherng2007,Rapp2007,chetcuti2021probe} to specific situations in condensed matter physics \cite{keller2014emergent,kugel2015spin,PhysRevLett.96.256602,arovas1999n}. The higher symmetry accounts for a variety of novel phenomena, such as symmetry-protected topological phases, and Mott-insulator transitions at finite interaction values \cite{Capponi2016a}. Recently, the research scope on SU($N$) fermions has been substantially enlarged by cold atom quantum technology \cite{cazalilla2014ultracold,Sowiski_2019}. Specifically, experiments with alkaline-earth and ytterbium atoms have simulated  SU($N$) fermions \cite{Pagano2014,PhysRevLett.113.120402,scazza2014observation} with $N$ as large as ten. SU($N$) cold atoms at the mesoscopic scale can provide a new platform for atomtronic circuits to widen the scope of current quantum simulation and access quantum devices with new specifications \cite{Amico2020}. In particular, by using the logic of current-voltage characteristics of solid state physics, an important goal of the atomtronics field is to eventually exploit matter-wave currents to probe quantum phases of matter. For SU($N$) fermions, such a program has been initiated by studying the persistent current \cite{Chetcuti2020a,andrea2021interaction}.

In this paper, we access the persistent current by means of a VQE. To this end, we generalize the Jordan-Wigner (JW) transformation applied to two spin-component fermions \cite{Shastry1986, Reiner_2016} to SU($N$) fermions. We apply our fermion-to-qubit mapping to SU($N$) Hubbard models in which a density-density repulsion is also considered, further to the characteristic local interaction. By careful analysis of the complexity and performance of the VQE, we demonstrate that the quantum correlations characterizing the ground-state of system can be captured only if quantum circuits with a suitable depth are considered. We shall see that SU($N$) fermionic systems require a larger number of layers with respect to two-component fermionic systems, and how the persistent current displays distinctive features in the different quantum phases of the system.

\section{SU(\textit{N}) Fermion-to-Qubit Mapping} \label{S.mapping}

Our physical system is made of a set of SU($N$) fermions localized in a 1D ring lattice with $L$ sites:
\begin{align}
\label{E.FH}
H = -t &\sum_{i=0}^{L-1}\sum_{s=0}^{N-1} \left(e^{\imath \frac{2 \pi \phi}{L}} c_{i, s}^\dagger c_{i+1,s} +\text{h.c.} \right) + U\sum_{i=0}^{L-1} \sum_{s=0}^{N-1}\sum_{s'=s+1}^{N-1} n_{i, s}n_{i, s'} + V\sum_{i=0}^{L-1} n_{i} n_{i+1} \,,
\end{align}
where $c_{i, s}^\dagger$ ($c_{i, s}$) creates (annihilates) a fermion with spin-component $s$ at site $i$. The $s$-fermion number operator is $n_{i, s} = c_{i, s}^\dagger c_{i, s}$, and the total number operator defined as $n_i = \sum_s n_{i, s}$ on-site $i$, with h.c. denoting the Hermitian conjugate of preceding terms. The parameter $t$ is the hopping amplitude of a fermion between nearest-neighbor sites, while the parameters $U$ and $V$ describe the on-site and density-density nearest-neighbor interaction, respectively. In this paper, only systems with an equal number of particles in each color are considered. Furthermore, we only consider non-negative values for both $U$ and $V$ (refer to \ref{general} for generalizing Eq. \ref{E.FH} to a higher-dimensional long-range Hubbard model).

For $V=0$, a superfluid to Mott insulator transition takes place for a finite value of the local repulsive interaction $U$ \cite{Capponi2016a,Xu2018}. Such a transition occurs at finite $U$ for $N>2$. Because of the interplay between $U$ and $V$, different quantum phases can be displayed. Specifically, the phase diagram involves a superfluid phase, a Mott phase, and a ``beat'' phase in which the particle occupation is modulated along the chain with a vanishing spin gap \cite{extendedphase}.

\begin{figure}[!ht]
	\centering
	\includegraphics[width=\textwidth]{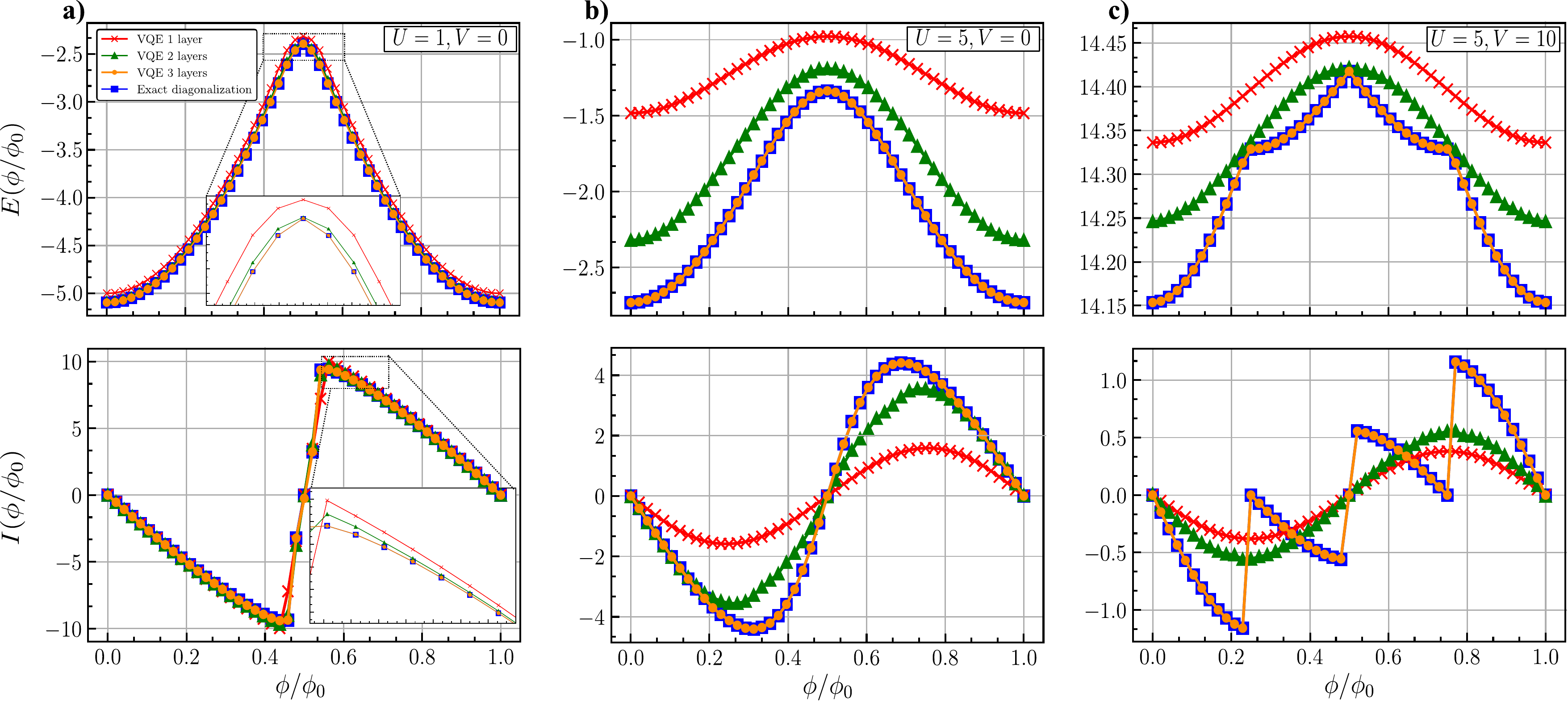}
	\caption{The ground-state energy $E_{0}(\phi)$ (top panel) and the corresponding persistent current $I(\phi)$ (bottom panel) for SU(3) fermions with different local $U$ and nearest-neighbor $V$ interactions, in the integer filling regime of the Hubbard model, using exact energy measurements. The profile of the persistent current gives a clear indication between the \textbf{a)} superfluid, \textbf{b)} Mott and \textbf{c)} beat phases. Exact diagonalization for $L=N_{p}=3$ is used to monitor the results obtained by the VQE reported in Fig. \ref{fig:circuit} (with a varying number of layers). Three layers are required in order for the ground-state energy of the latter to match the former.}
	\label{fig:3comm}
\end{figure}
 
The phase factor $e^{\imath \frac{2 \pi \phi}{L}}$ in Eq. \eqref{E.FH} takes into account the effective magnetic flux piercing the ring, which is able to impart a persistent current with the following form:
\begin{align}
\label{E_persis1}
I\left(\phi\right)=\frac{2\pi\imath t}{L}\sum\limits_{i=0}^{L-1}\sum\limits_{s=0}^{N-1}\bra{\psi_{0}}e^{\imath\frac{2\pi\phi}{L}}c_{i,s}^{\dagger}c_{i+1,s} -\mathrm{h.c}.\ket{\psi_{0}} \,,
\end{align}
where $\ket{\psi_{0}}$ denotes the ground-state of Eq. \eqref{E.FH}. For the case of $V=0$, it has  been recently demonstrated that as a combination of the interaction, magnetic flux and spin correlations, the effective elementary flux quantum $\phi_{0}$ which fixes the persistent current periodicity, is observed to evolve from a single particle one to an extreme case of fractional flux quantum, in which one flux quantum is shared by all the particles. Such a phenomenon reflects a type of attraction from repulsion: despite the repulsive interaction, spin correlations can lead to many-body states reacting as if they were bound states of fermions \cite{Chetcuti2020a}.

To implement the VQE on a NISQ computer, it is crucial to represent the system described by Eq. \eqref{E.FH} in terms of qubits \cite{McArdle2020,Cai2020, Dallaire-Demers2016,Cade2020}. To this end, we extend the JW transformation originally devised for two-component fermions \cite{1928JW,Shastry1986, Reiner_2016} to the general SU($N$) case as done, e.g., in Ref.~\cite{Roux2009a} for the mapping of SU(4) fermions to hard-core bosons.  We introduce $N$ sets of Pauli operators, one for every spin-component $s$, hereafter called color, of the fermionic atom. The mapping assumes the following form:
\begin{equation}
\label{E.map}
    c_{i,s}^\dagger=\prod_{j<n}\sigma_j^z\sigma_n^+ \,,\, c_{i,s}=\prod_{j<n}\sigma_j^z\sigma_n^- \,,\, (i, s) \xrightarrow{} n = i + sL \,,
\end{equation}
where $\sigma^\pm = (\sigma^x \pm \imath\sigma^y) / 2$ are the ladder operators, and $\ket{0}_n$ and $\ket{1}_n$ represent the absence and presence of a fermion of color $s$ on site $i$, respectively. In this way, up to the Pauli-string operator $\prod_{j<n}\sigma_j^z$, a fermion of color $s$ on-site $i$ is mapped onto a spin-$\frac{1}{2}$ operator $\sigma^+_n$ acting on qubit $n=i+sL$ (of the qubit lattice). As a consequence, a system of $N_p$ SU($N$) fermions on $L$ sites is mapped into $NL$ qubits. It is straightforward to show that the mapping in Eq. \eqref{E.map} preserves fermionic commutation rules independently of the chosen order of the color $s$ appearing in the definition of $n$. In the following, without loss of generality, we adopt an increasing order for $s$, meaning $n'>n$ for fermionic operators with $s'>s$. The Hamiltonian of Eq. \eqref{E.FH} is thus mapped to
\begin{align}
    H= &-t\sum\limits_{i, s} P_{i,s}\left(e^{\imath \frac{2 \pi \phi}{L}} \sigma_{i+sL}^+\sigma_{i+1+sL}^-+\text{h.c.}\right) \nonumber\\
    &+ \frac{U}{4}\sum\limits_{i, s < s'}\left( 1-\sigma_{i+sL}^z \right)\left( 1-\sigma_{i+s'L}^z \right)\nonumber \\
    &+ \frac{V}{4}\sum\limits_{i, s, s'} \left( 1-\sigma_{i+sL}^z \right)\left( 1-\sigma_{i+1+s'L}^z \right) \,,
\label{eq:map1}
\end{align}
where $P_{i,s}$ is the color-dependent parity term given by
\begin{equation}
    P_{i,s} = \begin{cases}
        -1 \,,\, \text{if } i = L - 1 \text{ and } N_s \text{ is odd,} \\
        +1 \,,\, \text{otherwise,}
        \end{cases}
\end{equation}
with $N_s$ being the number of fermions of color $s$.

The aim of the VQE is to find a set of parameters $\Vec{\theta}$ able to obtain a quantum state $\ket{\psi(\Vec{\theta})}$, that minimizes the expectation value of the energy $E(\Vec{\theta})$ of a Hamiltonian, through a shallow parameterized quantum circuit $U(\Vec{\theta})$ acting on an initial state $\ket{\psi}$. The optimization is achieved through an adiabatically-assisted quantum-classical loop of a classical minimization process using evaluations of the parameterized quantum circuit \cite{AAVQE}. We propose a variational ansatz utilizing the extension of the JW transformation to SU($N$) fermions, inspired from Ref. \cite{Cade2020} which adopted a number-preserving ansatz for SU(2) fermions. Parameterized $\imath$SWAP gates model the hopping terms between sites representing the same color, while the interaction terms, as well as the on-site terms, between fermions of a different color, are represented by controlled-$R_z$ and $R_z$ gates, respectively.

We have simulated the quantum circuits using the {\tt Qibo} API \cite{efthymiou2020qibo, stavros2020qibozenodo}, disregarding the noise of the quantum gates. The main text contains only simulations with exact energy measurements, while in \ref{S.sampling} there are results highlighting the effects of finite sampling in the optimization procedure. The classical optimization technique employed in the quantum-classical loop, in the case of statevector simulations (exact energy measurements), is the BFGS method \cite{nocedal2006bfgs}, a gradient-based approach which involves the computation of the inverse Hessian matrix. Finally, we benchmark our results using the exact diagonalization methods provided by the {\tt QuSpin} software package \cite{Weinberg2016, Weinberg2019a}. In our approach, we monitor the  persistent current given in Eq. \eqref{E_persis1}, that we access by means of a VQE subroutine. Details on the VQE implementation can be found in \ref{S.VQE}.

\section{Results} \label{S.results}

\begin{figure}[!t]
    \centering
    \includegraphics[width=0.7\textwidth]{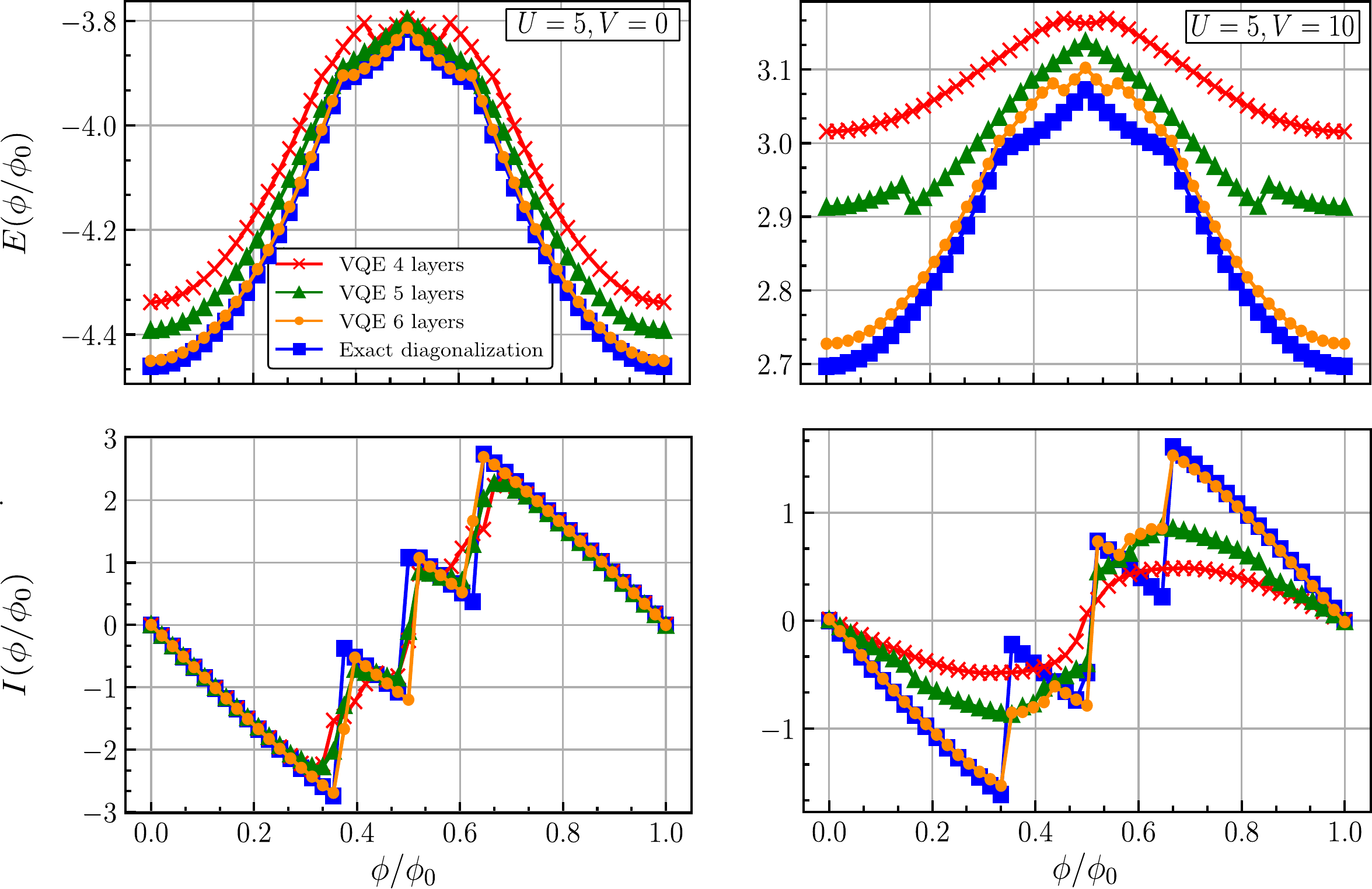}
    \caption{The ground-state energy $E_{0}(\phi)$ (top panel) and the corresponding persistent current $I(\phi)$ (bottom panel) for SU(3) fermions with different local $U$ and nearest-neighbor $V$ interactions, in the incommensurate regime of the Hubbard model, using exact energy measurements. Exact diagonalization for $L=5$ and $N_{p}=3$ is used to monitor the results obtained by VQE.}
    \label{fig:incommensurate}
\end{figure}

First, we discuss the results of the VQE for commensurate and incommensurate SU(3) models found in Figs. \ref{fig:3comm} and \ref{fig:incommensurate}, respectively, with different values of the parameters $U$ and $V$ corresponding to different physical regimes in the phase diagram of the system.

In the lower panels of Figs. \ref{fig:3comm} and \ref{fig:incommensurate} we monitor the persistent current given in Eq. \eqref{E_persis1}. In the commensurate case, we observe that the persistent current has a sawtooth shape as in Fig.~\ref{fig:3comm} \textbf{a)}. Upon increasing the on-site interaction, the sawtooth shape is smoothed out into a sinusoidal one due to the opening of the spectral gap, which indicates the onset to the Mott phase transition~\cite{Chetcuti2020a} --- Fig.~\ref{fig:3comm} \textbf{b)}. The last panel of Fig.~\ref{fig:3comm} depicts a very interesting phenomenon. By increasing nearest-neighbor interactions through $V$, one goes from the Mott to the beat phase~\cite{extendedphase}, where the latter is characterized by the modulation of the density arising out of effective attraction from repulsion. This in turn results in a hybrid form of the persistent current: a smoothed out yet fractionalized one. Figs.~\ref{fig:incommensurate} \textbf{a)} and \textbf{b)} display the persistent current against interaction for incommensurate systems. Upon increasing $U$, the persistent current fractionalizes due to energy level crossings that occur to counteract the increase in $\phi$, resulting in the creation of spinon excitations in the ground-state~\cite{Chetcuti2020a}. Upon increasing $V$, as in the lower panel of Fig.~\ref{fig:incommensurate} \textbf{b)}, the fractionalization is enhanced due to the increased repulsion in the system.

At $\phi = 0$, the final optimal parameters of each case were determined by training the VQE with the random initialization of parameters. However, each subsequent instance of the VQE, tasked with finding the ground-state energy of the model with the next iteration of $\phi$, was fed the optimal parameters from the previous iteration. This form of adiabatic assistance, along with the symmetry of the ground-state energy along the degeneracy point $\phi/\phi_{0} = 0.5$, offers a significant speed up in mapping out the ground-states of the extended Hubbard Hamiltonian over the entire range of $\phi$, which is necessary for determining the persistent current.

We notice that the discontinuities of $I\left(\phi\right)$, due to spinon creation in the ground-state, are fully captured. We note that such a phenomenon is only captured by the VQE if a sufficient number of layers are considered. This feature is consistent with the general theory signifying that the flux quantum fractionalization is a genuine many-body correlation effect. We also notice that, for small $U$ and $V$, a shallow circuit estimates the ground-state with high accuracy. However, the correlations that are present for larger $U$ and $V$ necessitate the need for more ansatz layers to fully capture the ground-state of the model. We use the entanglement entropy of the final state as an additional measure to gauge how closely the VQE reproduces the desired ground-state.

\begin{figure}[!t]
    \centering
    \includegraphics[width=0.7\textwidth]{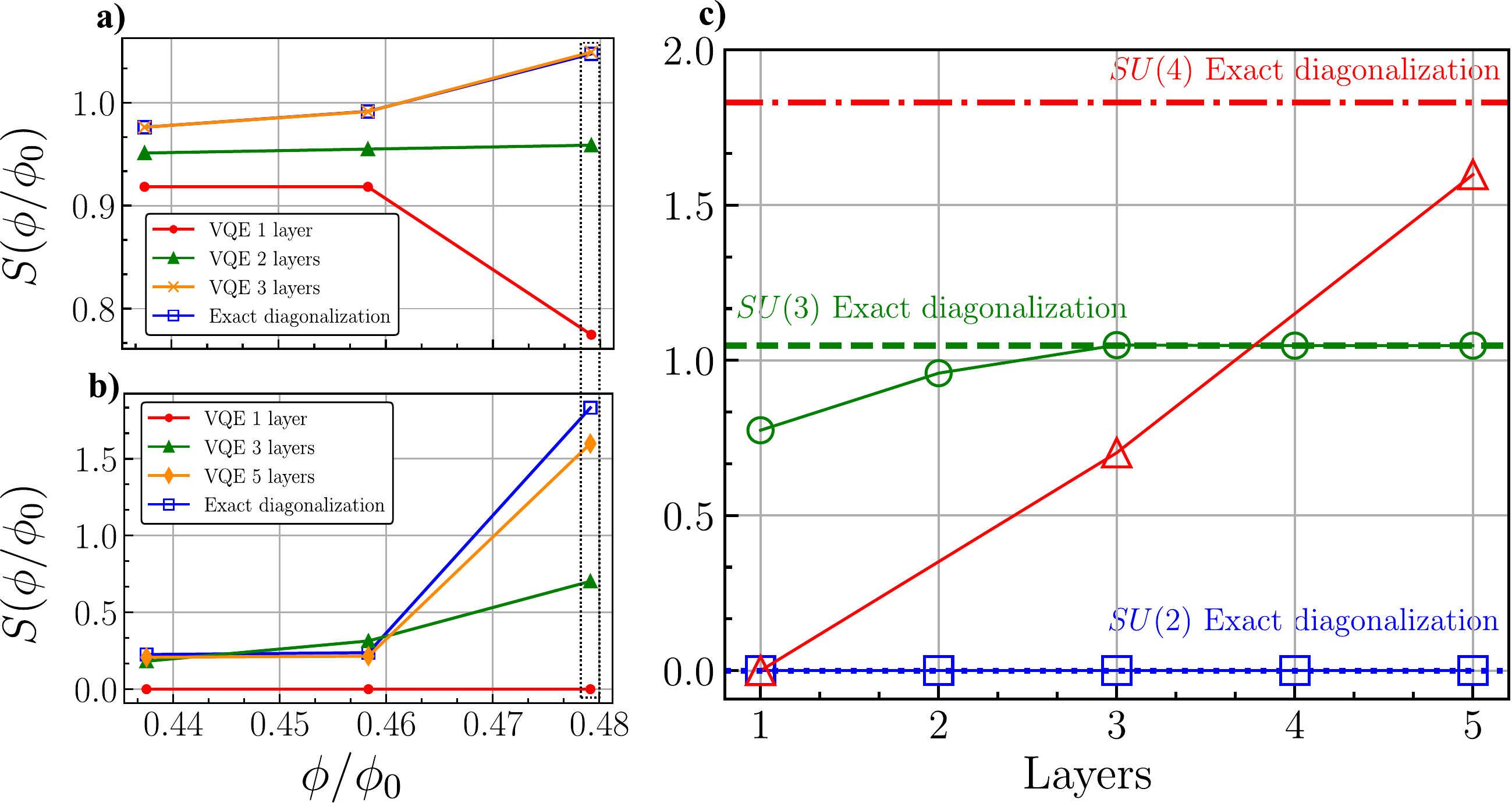}
    \caption{Values for the bipartite entropy $S(\phi)$ of half the qubit chain for \textbf{a)} SU(3) and \textbf{b)} SU(4) for an increasing number of ansatz layers for a Hubbard model with $U=1$ and $V=0.5$ at commensurate fillings, namely, $N_p=N$, using exact energy measurements. \textbf{c)} By looking specifically at $\phi=0.5$, the SU(2) and SU(3) cases quickly reach the entropy of the exact diagonalization state, while the SU(4) case still requires more correlations to be built up within the quantum circuit to fully approximate the ground-state (see Fig. \ref{fig:su4} in \ref{S.SU4}).}
    \label{fig:entanglement}
\end{figure}

We note that, by increasing the number of layers of the parameterized quantum circuit, the entanglement in the ground-state increases, as shown in Fig. \ref{fig:entanglement}. This layer-by-layer build up of entanglement is crucial to capture the correlations of the target quantum state. It is only after the parameterized quantum circuit has enough depth to reach the target entanglement entropy, that the VQE can start to approximate the target observable with high precision, as previously shown in Ref. \cite{BravoPrieto2020}. We observe that correctly taking the ground-state entanglement into account is very important to capture the physics of the system.

The number-preserving ansatz we introduce explores a subspace of the Hilbert space of size $\prod_{s=0}^{N-1}\binom{L}{N_s}$ for SU($N$) fermions in $L$ sites, where $N_s$ is the number of fermions of color $s$. Therefore, the number of layers needed to properly capture the ground-state energy will depend on the expressibility of this ansatz in covering the reduced subspace. In order to reduce the number of parameters needed for the VQE, one could introduce a translationally invariant ansatz, to account for the periodic nature of the fermionic chains studied. If, as in our examined models (Figs. \ref{fig:3comm} and \ref{fig:incommensurate}), we suppose that $N_s = N_{s'}$ for all colors $s$ and $s'$, then this symmetry reduces the relevant basis of the problem to $L^{N-1}$ states. However, this ansatz would need non-trivial and non-local quantum gates, which are not suitable for implementation in NISQ computers.

Further results showcasing the performance of the VQE on SU(4) commensurate models using exact energy measurements, and simulations employing finite sampling for the commensurate SU(3) case, as in Fig. \ref{fig:3comm}, are presented in \ref{sec:further_results}.

\section{Conclusion} \label{S.Concl}

In this work, we simulated a VQE program on SU($N$) fermions. As an important technical step, we extended the Jordan-Wigner transformation to $N$-component fermions. Our work provides a specific instance of a current-based quantum simulator: the physics of the system is probed by the current response to the gauge field. Specifically, we consider a 1D ring lattice of the Hubbard type, and we monitor the persistent current.

We have then applied the SU($N$) fermion-to-qubit mapping, using it to determine the ground-state energy of the Hubbard Hamiltonian via the VQE. We have found that the ground-state energy can be approached by utilizing a number-preserving ansatz, similarly to what is done for the SU(2) case \cite{Cade2020}. However our ansatz is generalized to the SU($N$) fermion case, and optimized to minimize the depth, and the number of gates and parameters needed to achieve the ground-state of the magnetic-flux-induced extended Hubbard Hamiltonian. In this context, it would be interesting to consider SU($N$) generalization of other two-component fermion-to-qubit mappings, such as the Bravyi-Kitaev \cite{Bravyi2002} and Ball-Verstraete-Cirac \cite{Ball2005,Verstraete2005} encodings. Further optimization to this model would include the so-called state preparation and measurement mitigation \cite{sun2020efficient}, contextual optimization \cite{context}, implementation of exchange symmetries \cite{decamp2016exact,nataf2016exact}, noise tailoring through randomization protocols \cite{Wallman_2016}, quantum machine learning meta-techniques \cite{cevera-lierta2021}, and noise-aware logical-to-physical qubit mappings \cite{White_2021} when running on quantum hardware, see Ref.~\cite{Tilly2021} for a recent review on VQE. These methods for minimizing noise could be investigated and applied to the proposed parameterized quantum circuit, in order to obtain a further optimized algorithm for the determination of the ground-state energy of the Hubbard Hamiltonian (and other similar quantum models) on NISQ computers.

In view of the recent interest for the SU($N$) Hubbard model, motivated both by its experimental realization with cold atoms, and its relevance in modeling mesoscopic quantum devices stemming from the rapidly developing atomtronics technology, our work opens this model for investigation via variational quantum algorithms. 

Unlike the SU($N$) Hubbard model, which becomes Bethe ansatz integrable under certain parameters and filling fractions, the extended version is not, that is the Hubbard model with $V \neq 0$. One must rely on numerical methods such as exact diagonalization. However, this technique is only feasible for small systems, as the Hilbert space increases exponentially in both the system size, and number of spin-components $N$. Other methods such as DMRG, which is regarded as being optimal for one-dimensional systems, also face limitations due to scalability~\cite{LeBlanc2015}. As a result, one is always looking for the development of new methods, capable of efficient numerical computation which scale in sub-exponential time and memory. The VQE is one such alternative. It is believed that the use of a quantum system to compute the ground-state of Hamiltonians may help overcome the exponential scaling that comes with classical simulation~\cite{mcclean2016theory}.

At the same time, the proposed SU($N$) fermion-to-qubit mapping may prove useful in addressing the Hubbard model via classical algorithms originally developed for SU(2) models, such as tensor network states \cite{Orus2019}, and specifically, PEPS \cite{Bruognolo2021}, providing further incentive in investigating some of the results obtained in this paper. The {\tt Python} code used in obtaining the results is also publicly available on {\tt GitHub} \cite{qibo_code}.

\section*{Acknowledgements}

We acknowledge fruitful discussions with Tobias Haug.

\appendix

\section{Further Results} \label{sec:further_results}

In this section we present further results obtained from taking sampled energy measurements, rather than exact ones, for the case of SU(3) as in Fig. \ref{fig:3comm}. We also show results of statevector simulations for SU(4) Hubbard systems.

\subsection{Statevector simulations for SU(4)}\label{S.SU4}

\begin{figure}[ht]
    \centering
    \includegraphics[width=0.7\textwidth]{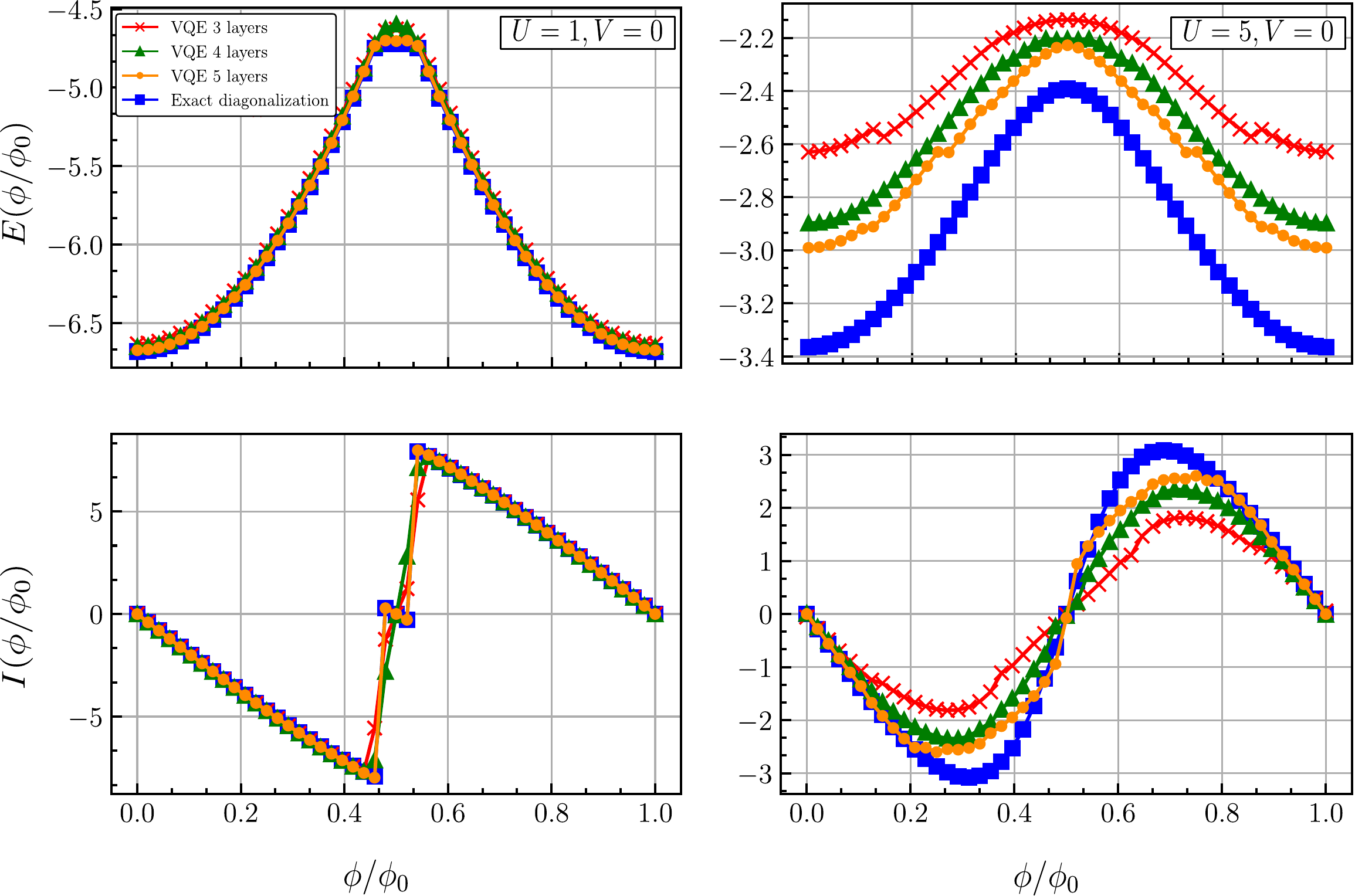}
    \caption{The ground-state energy $E_{0}(\phi)$ (top panel) and the corresponding persistent current $I(\phi)$ (bottom panel) for SU(4) fermions with different local $U$ and nearest-neighbor $V$ interactions, in the integer filling regime of the Hubbard model, using exact energy measurements. Exact diagonalization for $L=N_{p}=4$ is used to monitor the results obtained by VQE.}
    \label{fig:su4}
\end{figure}

Here we showcase the results with matching parameters ($U$ and $V$) as in Fig. \ref{fig:3comm}, but for the case of a commensurate SU(4) Hubbard model. In Fig. \ref{fig:su4}, for small values of $U$ and $V$ (left panels), five layers are enough to capture the ground-state energy, yet for higher interaction terms (right panels) the VQE does not build up enough correlations in order to approximate the ground-state energy (see Fig. \ref{fig:entanglement}).

\subsection{Optimization with Sampling} \label{S.sampling}

Here we present the results of taking a finite number of samples during the optimization procedure. For the following simulations, after testing the performance of several optimizers available at Ref. \cite{Qiskit_optimizers}, we chose the NFT optimizer \cite{NFT}. The reason being that this optimizer provided the best results, due to its consistency and analytical gradient-free optimization technique, as recommended in Ref. \cite{Tilly2021}.

\begin{figure}[ht]
    \centering
    \includegraphics[width=1\textwidth]{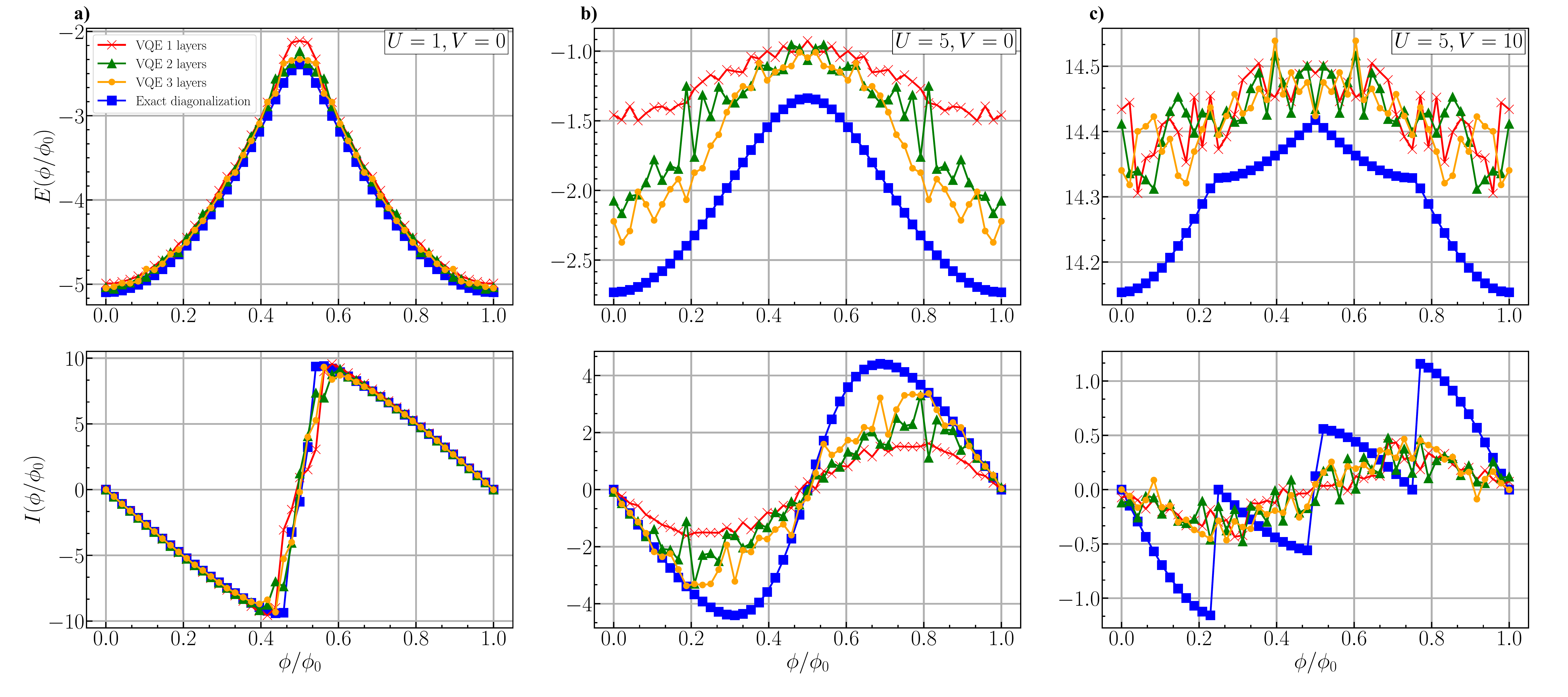}
    \caption{The ground-state energy $E_{0}(\phi)$ (top panel) and the corresponding persistent current $I(\phi)$ (bottom panel) for SU(3) fermions with different local $U$ and nearest-neighbor $V$ interactions, in the integer filling regime of the Hubbard model, using sampled energy measurements. The profile of the persistent current gives a clear indication between the \textbf{a)} superfluid, \textbf{b)} Mott and \textbf{c)} beat phases. Exact diagonalization for $L=N_{p}=3$ is used to monitor the results obtained by the VQE reported in Fig. \ref{fig:circuit} (with a varying number of layers).}
    \label{fig:sampled_su3}
\end{figure}

In Fig. \ref{fig:sampled_su3}, a total of 65536 function evaluations were carried out for each case. For small values of $U$ and $V$, similar to the results of exact energy measurements, even just one layer is sufficient to nearly capture the ground-state energies and corresponding persistent current, with three layers acquiring a close-to-perfect accuracy. However, in contrast with the statevector simulations, three layers are not enough to capture the ground-state energies for higher values of $U$ and $V$. Running a VQE using sampled energy measurements is undeniably more complex, with optimizers typically more prone to getting stuck in a barren plateau or local minima, as well as taking significantly longer to converge.

\begin{figure}[ht]
    \centering
    \includegraphics[width=\textwidth]{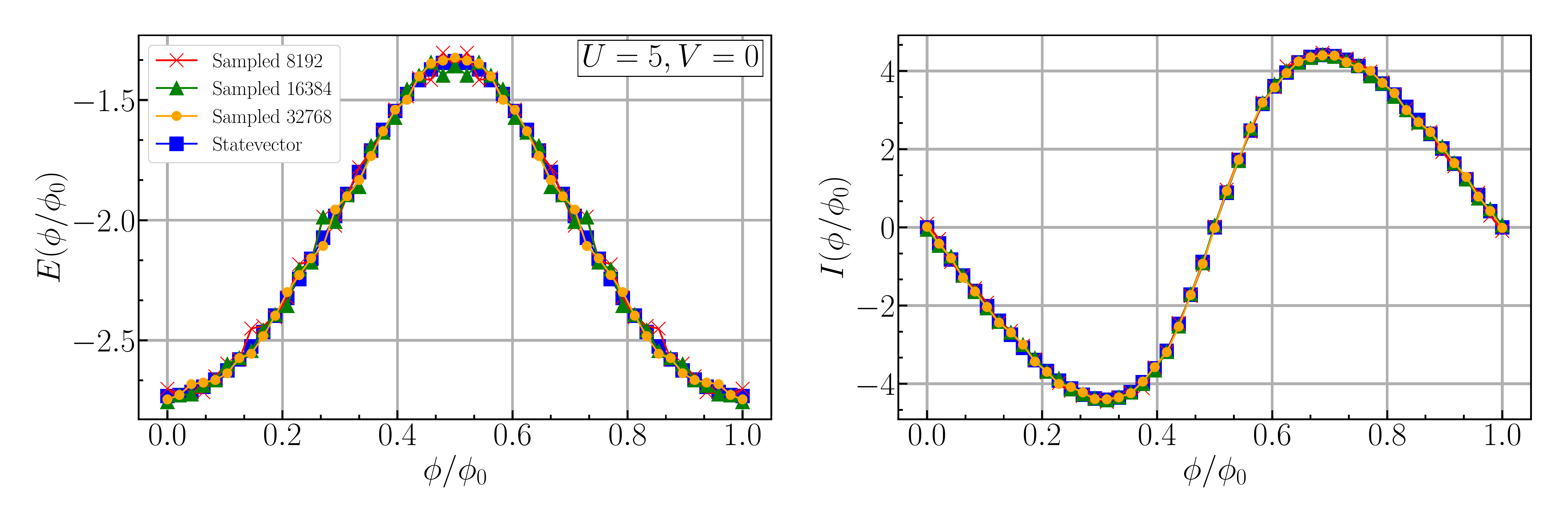}
    \caption{The ground-state energy $E_{0}(\phi)$ (left panel) and the corresponding persistent current $I(\phi)$ (right panel) for SU(3) fermions with $U=5$ and no nearest-neighbor interaction, in the integer filling regime of the Hubbard model. The statevector results were obtained with exact energy measurements. On the other hand the sampled profiles were obtained by reevaulating the ground-state energies using the parameters obtained via the statevector results.}
    \label{fig:sampled_statevector}
\end{figure}

To ascertain that the sampling itself was not the problem, we used the statevector parameters of the ground-states of Fig. \ref{fig:3comm} b) ($U = 5, V = 0$) to reevaluate the same ground-state energies, but instead using finite sampling. The results are shown in Fig. \ref{fig:sampled_statevector}. With just 8192 samples, the ground-state energies and corresponding persistent current are obtained with sufficient accuracy, with 16384 and 32768 samples converging even more so to the statevector results. It is important to note that the statistical error due to finite sampling is in the order of $1/\sqrt{n}$, where $n$ is the number of samples.

\begin{figure}[ht]
    \centering
    \includegraphics[width=\textwidth]{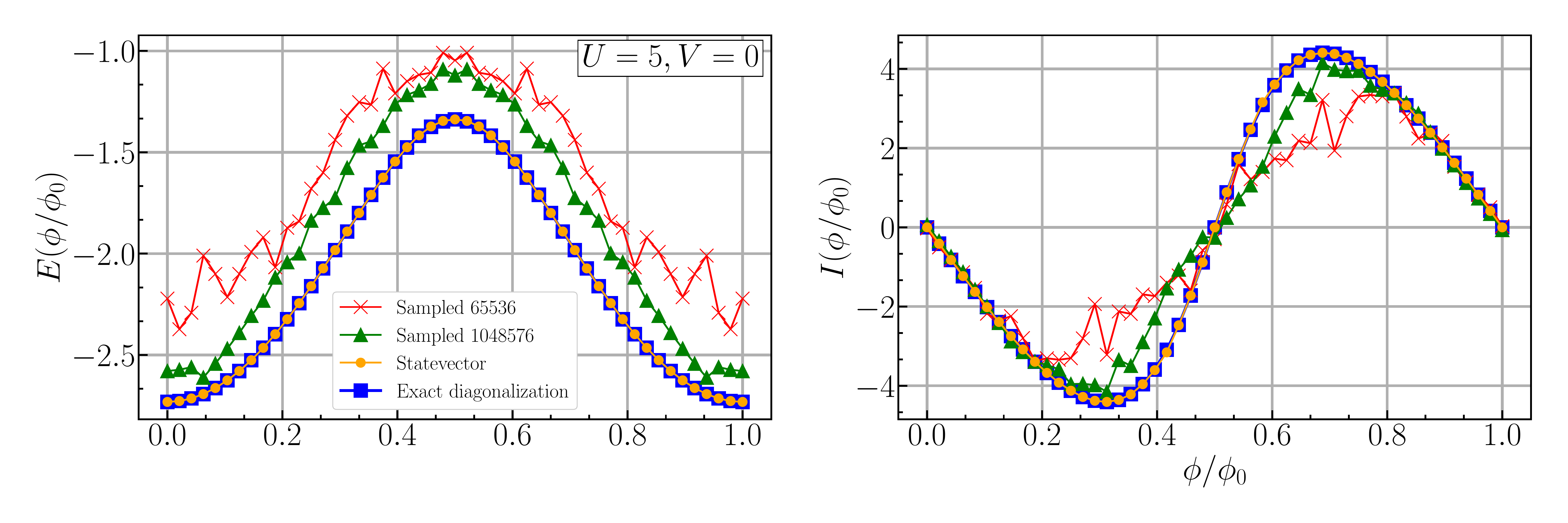}
    \caption{The ground-state energy $E_{0}(\phi)$ (left panel) and the corresponding persistent current $I(\phi)$ (right panel) for SU(3) fermions with $U=5$ and no nearest-neighbor interaction, in the integer filling regime of the Hubbard model. The statevector results were obtained with exact energy measurements. On the other hand the sampled profiles (each 32678 samples) were obtained by carrying out the optimization procedure with a different number of function evaluations in the NFT optimizer. Exact diagonalization for $L=N_{p}=3$ is used to monitor the results obtained by the VQE reported in Fig. \ref{fig:circuit}.}
    \label{fig:sampled_fevs}
\end{figure}

To justify our argument that the optimization procedure needs improvement, we have also simulated the case of Fig. \ref{fig:3comm} b) ($U = 5, V = 0$) with three layers, using 1048576 function evaluations, and comparing it with the result of the one with 65536 function evaluations (Fig. \ref{fig:sampled_fevs}). We obtain a better representation of the ground-state energies using a higher number of function evaluations, however, although close, the energies still do not match the statevector simulations, with deviations in the range between $3\%$ and $19\%$, (left panel of Fig. \ref{fig:sampled_fevs}). Interestingly, the curvature of the energy values as a function of $\phi$, as represented by the persistent current, is well captured even with sampling in a wide interval of $\phi$, (right panel of Fig. \ref{fig:sampled_fevs}). This suggests that more work needs to be devoted to improving the optimization procedure when utilizing finite sampling instead of exact energy measurements.

\section{Details on the Qubit Hamiltonian}\label{general}

The details of the SU($N$) magnetic-flux-induced Hubbard Hamiltonian shown in Eq. \eqref{E.FH} are defined and expanded here. First we show how we obtained the qubit Hamiltonian from the SU($N$) JW transformation, followed by extending the Hamiltonian to cater for both long-range hopping and interaction terms.

\subsection{Obtaining the qubit Hamiltonian via the SU(\textit{N}) JW Transformation}

The original Hubbard Hamiltonian (Eq. \eqref{E.FH}) is
\begin{align}
    H = -t\sum_{i, s}\left(e^{\imath \frac{2 \pi \phi}{L}} c_{i, s}^\dagger c_{i+1,s} +\text{h.c.} \right) + U\sum_{i, s < s'}n_{i, s}n_{i, s'} + V\sum_{i}n_{i} n_{i+1} \,,
    \label{eq:Ham}
\end{align}
with the SU($N$) JW transformation being
\begin{equation}
    c_{i,s}^\dagger=\prod_{j<n}\sigma_j^z\sigma_n^+ \,,\, c_{i,s}=\prod_{j<n}\sigma_j^z\sigma_n^- \,,\, (i, s) \xrightarrow{} n = i + sL \,,
    \label{eq:JW}
\end{equation}
Starting with some preliminary calculations, notice that $\sigma^z_n \sigma^\pm_n = \mp \sigma^\pm_n$ and $\sigma^\pm_n \sigma^z_n = \pm \sigma^\pm_n$. Looking at only the nearest-neighbor hopping terms, we find that
\begin{align}
    c^\dagger_{i, s}c_{i+1, s} 
    &= \left(\prod_{j<n}\sigma_j^z\sigma_n^+\right) \left(\prod_{k<n+1}\sigma_k^z\sigma_{n+1}^-\right) \nonumber \\ 
    &= \sigma^+_n\sigma^z_n\sigma^-_{n+1} \nonumber \\ 
    &= \sigma^+_n\sigma^-_{n+1} \,.
\end{align}
Similarly, looking at the Hermitian conjugate terms,
\begin{align}
    c^\dagger_{i+1, s}c_{i, s} 
    &= \left(\prod_{j<n+1}\sigma_j^z\sigma_{n+1}^+\right) \left(\prod_{k<n}\sigma_k^z\sigma_n^-\right) \nonumber \\ 
    &= \sigma^+_{n+1}\sigma^z_n\sigma^-_{n} \nonumber \\ 
    &= \sigma^-_n\sigma^+_{n+1} \,.
\end{align} 
Note that if $i = L - 1$, then $i + 1 = 0 \mod L$, meaning that
\begin{align}
    c^\dagger_{L-1, s}c_{0, s} 
    &= \left(\prod_{j<L-1+sL}\sigma_j^z\sigma_{L-1+sL}^+\right) \left(\prod_{k<sL}\sigma_k^z\sigma_{sL}^-\right) \nonumber \\
    &= \sigma^-_{sL}\left(\sum_{j=sL}^{L-1+sL}\sigma^z_j\right)\sigma^+_{L-1+sL} \nonumber \\
    &= \sigma^-_{sL}\left(\sum_{j=sL}^{L-1+sL}\sigma^z_j\right)\sigma^+_{L-1+sL} \nonumber \\
    &= \sigma^-_{sL}\left(\sum_{j=1+sL}^{L-2+sL}\sigma^z_j\right)\sigma^+_{L-1+sL} \,,
\end{align}
and similarly for the Hermitian conjugate terms. This results in a parity term, which is $-1$ at the ``looping'' index $i=L-1$ only when the number of spins of color $s$, $N_s$, is odd, otherwise it is $+1$. Thus in the Hamiltonian we can replace the $\sigma^z$ terms with
\begin{equation}
    P_{i,s} = \begin{cases}
        -1 \,,\, \text{if } i = L - 1 \text{ and } N_s \text{ is odd,} \\
        +1 \,,\, \text{otherwise.}
        \end{cases}
\end{equation}
Now, since $n_{i, s} = c^\dagger_{i, s}c_{i, s}$, then
\begin{equation}
    n_{i, s} = \left(\prod_{j<n}\sigma_j^z\sigma_n^+\right) \left(\prod_{k<n}\sigma_k^z\sigma_n^-\right) = \sigma^+_n \sigma^-_n \,,
\end{equation}
which can be rewritten as
\begin{equation}
    n_{i, s} = \dfrac{1 - \sigma^z_n}{2} \,.
\end{equation}
Therefore, looking at the interaction terms of $U$, we have
\begin{equation}
    n_{i, s}n_{i, s'} = \left(\dfrac{1 - \sigma^z_{i + sL}}{2}\right)\left(\dfrac{1 - \sigma^z_{i + s'L}}{2}\right) \,,
\end{equation}
while for the interaction terms of $V$,
\begin{align}
    n_i n_{i + 1} 
    &= \left(\sum_s n_{i, s}\right) \left(\sum_{s'} n_{i+1, s'}\right) \nonumber \\
    &= \sum_{s, s'} \left(\dfrac{1 - \sigma^z_{i + sL}}{2}\right)\left(\dfrac{1 - \sigma^z_{i + 1 + s'L}}{2}\right) \,.
\end{align}
Thus, starting from our original fermionic Hamiltonian \eqref{eq:Ham}, and applying the SU($N$) JW transformation \eqref{eq:JW}, we end up with the following representation:
\begin{align}
    H= &-t\sum\limits_{i, s} P_{i,s}\left(e^{\imath \frac{2 \pi \phi}{L}} \sigma_{i+sL}^+\sigma_{i+1+sL}^-+ \text{h.c.}\right) \nonumber\\
    &+ \frac{U}{4}\sum\limits_{i, s < s'}\left( 1-\sigma_{i+sL}^z \right)\left( 1-\sigma_{i+s'L}^z \right) \nonumber \\
    &+ \frac{V}{4}\sum\limits_{i, s, s'} \left( 1-\sigma_{i+sL}^z \right)\left( 1-\sigma_{i+1+s'L}^z \right) \,,
\end{align}
which is exactly the derived qubit Hamiltonian (Eq. \eqref{eq:map1}) in the main manuscript.

\subsection{Extending the Hamiltonian}

The Hubbard Hamiltonian in Eq. \eqref{E.FH} can be generalized to incorporate symmetric long-range hopping and interaction terms, transforming into 
\begin{align}
\label{E.FH1}
 H = -&\sum_{i=0}^{L-1} \sum_{s=0}^{N-1} \sum_{r=1}^{R_t} t_{r} \left(e^{\imath \frac{2 \pi \phi}{L}}  c_{i, s}^\dagger c_{i+r,s} +\text{h.c.} \right) \nonumber \\ 
 +&\ U\sum_{i=0}^{L-1}\sum_{s=0}^{N-1}\sum_{s'=s+1}^{N-1} n_{i, s}n_{i, s'} \nonumber \\ 
 +&\ \sum_{i=0}^{L-1}\sum_{r=1}^{R_V} V_{r} n_{i} n_{i+r} \,,
\end{align}
where $t_{r}$ is the hopping amplitude of a fermion between sites at a distance $r$, $U$ is the on-site interaction, and $V_{r}$ is the interaction between fermions at a distance $r$. $R_V$ and $R_t$ represent the range of the interacting $V$ and hopping $t$ terms, respectively, which can be at most $\left\lfloor \frac{L}{2} \right\rfloor$, due to the periodic boundary conditions.

Given that the SU($N$) fermion-to-qubit mapping is independent of the parameters of the Hamiltonian, Eq. \eqref{E.FH1} (containing both site-dependent long-range hopping and interaction terms) can also be mapped to a qubit Hamiltonian, although with different geometry and interaction patterns:
\begin{align}
    H = &-\sum\limits_{i, s, r} t_r \left(e^{\imath \frac{2 \pi \phi}{L}} \sigma_{i+sL}^+\sigma_{i+r+sL}^-+ \text{h.c.} \right) \prod_{j=i+1+sL}^{i+r-1+sL} \sigma^z_j \nonumber \\
    &+ \frac{U}{4}\sum\limits_{i, s < s'}\left( 1-\sigma_{i+sL}^z \right)\left( 1-\sigma_{i+s'L}^z \right)\nonumber\\
    &+ \sum\limits_{i,r, s, s'} \frac{V_{r}}{4} \left( 1-\sigma_{i+sL}^z \right)\left( 1-\sigma_{i+r+s'L}^z \right) \,.
\label{eq:mapr}
\end{align}

\section{VQE Implementation}\label{S.VQE}

Determining the ground-state energy of the Hubbard model model requires the generation of an initial state, and a parameterized quantum circuit able to modify the quantum state, with observable measurements efficiently calculating the expectation value of the ground-state. As the number of fermions for each color is conserved, the spin configuration of the initial state, i.e. the number of qubits in the $\ket{1}$ state, must be preserved with regards to the mapping. As a consequence, the quantum gates (unitary operations) in the parameterized quantum circuit must be in the form of Pauli terms present in the Hubbard Hamiltonian.

Note that the ladder operators $\sigma^+$, $\sigma^-$, of the transformed hopping terms can be rewritten in terms of $\sigma^x$ and $\sigma^y$ Pauli operators (ignoring qubit indices and $\sigma^z$ operators without loss of generality) as follows:
\begin{align}
\label{eq:hopping}
	&\left(e^{\imath\frac{2\pi\phi}{L}}\sigma^+\sigma^- + e^{-\imath \frac{2 \pi \phi}{L}}\sigma^-\sigma^+\right) \nonumber \\ &= e^{\imath \frac{2 \pi \phi}{L}}\frac{(\sigma^x+\imath \sigma^y)}{2}\frac{(\sigma^x-\imath \sigma^y)}{2} + e^{-\imath \frac{2 \pi \phi}{L}}\frac{(\sigma^x-\imath \sigma^y)}{2}\frac{(\sigma^x+\imath \sigma^y)}{2} \nonumber \\ &= \frac{1}{4}\left( e^{\imath\frac{2\pi\phi}{L}} \left(\sigma^x\sigma^x - \imath \sigma^x\sigma^y + \imath \sigma^y\sigma^x + \sigma^y\sigma^y\right) + e^{-\imath\frac{2\pi\phi}{L}}\left(\sigma^x\sigma^x + \imath \sigma^x\sigma^y - \imath \sigma^y\sigma^x + \sigma^y\sigma^y\right) \right) \nonumber \\ &= \frac{1}{4} \left( \left(e^{\imath\frac{2\pi\phi}{L}} + e^{-\imath\frac{2\pi\phi}{L}}\right)\left(\sigma^x\sigma^x + \sigma^y\sigma^y\right) - \imath\left(e^{\imath\frac{2\pi\phi}{L}} - e^{-\imath\frac{2\pi\phi}{L}}\right)\left(\sigma^x\sigma^y - \sigma^y\sigma^x\right) \right) \nonumber \\ &= \frac{1}{2}\left(\cos\left(\frac{2\pi\phi}{L}\right)\left(\sigma^x\sigma^x + \sigma^y\sigma^y\right) + \sin\left(\frac{2\pi\phi}{L}\right)\left(\sigma^x\sigma^y - \sigma^y\sigma^x\right) \right) \,.
\end{align}
We take $(\sigma^x\sigma^x + \sigma^y\sigma^y)/2$ exponential operators to represent the sublayer consisting of hopping terms, such that
\begin{equation}
        e^{-\imath\frac{\theta}{2}\left(\sigma^x\sigma^x + \sigma^y\sigma^y\right)} =
        \left( \begin{array}{cccc}
        1 & 0 & 0 & 0 \\
        0 & \cos(\theta) & -\imath\sin(\theta) & 0 \\
        0 & -\imath\sin(\theta) & \cos(\theta) & 0 \\
        0 & 0 & 0 & 1
        \end{array} \right) =
    \vcenter{\hbox{\includegraphics[width=0.17\textwidth]{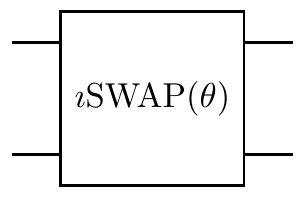}}}
\end{equation}
Similarly, for the interaction terms of $V$, we directly take the $(1-\sigma^z)(1-\sigma^z)/4$ terms in exponential form,
\begin{equation}
    e^{\imath\frac{\theta}{4}\left(1-\sigma^z\right)\left(1-\sigma^z\right)} =
    \left( \begin{array}{cccc}
    1 & 0 & 0 & 0 \\
    0 & 1 & 0 & 0 \\
    0 & 0 & e^{-\imath\theta} & 0 \\
    0 & 0 & 0 & e^{\imath\theta}
    \end{array} \right) =
    \vcenter{\hbox{\includegraphics[width=0.13\textwidth]{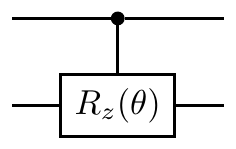}}}    
\end{equation}
Finally, for the on-site interaction terms representing $U$, we simply take the exponential terms of $(1-\sigma^z)/2$ to represent the sublayer, so that
\begin{equation}
    e^{\imath\frac{\theta}{2}\left(1-\sigma^z\right)} =
    \left( \begin{array}{cc}
		e^{-\imath\theta} & 0 \\
		0 & e^{\imath\theta}
	\end{array} \right) = \vcenter{\hbox{\includegraphics[width=0.13\textwidth]{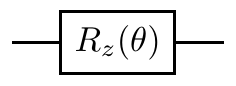}}}
\end{equation}
Now, to generate a specific spin configuration, $X$ gates can be used as an initial state, to create spins at particular sites with specified colors. In the simulations, created spins were followed by an initial sublayer of parameterized $R_z$ gates (which were omitted where no spin was created since they would not affect the overall quantum state).

All of the proposed quantum gates are number-preserving, which is crucial for ascertaining that the VQE searches for the ground-state of the SU($N$) Hubbard model in a specific spin configuration. An assembled circuit for a 3-site SU(3) nearest-neighbor Hubbard model is presented in Fig. \ref{fig:circuit}. This makes the variational ansatz more complex when compared with hardware-efficient ones. However, a symmetry-preserving ansatz provides certain advantages, such as potentially speeding up optimization and reducing the risk of barren plateaus, due to the restriction in the size of the Hilbert space \cite{Tilly2021}.

\subsection{Circuit Complexity}

Here we give a brief analysis of basis gate decompositions of the VQE ansatz of Fig. \ref{fig:circuit}, where we assume that the native gate set is composed of any one-qubit gate and controlled-NOT (CNOT) gates, and having full qubit connectivity. For the sake of simplicity, in the following we determine the effective circuit depth of a single ansatz layer by counting CNOT gates, and assuming that the gate times and errors of one-qubit gates are significantly less than those of a typical CNOT gate \cite{Willsch_2017}. A single ansatz layer consists of an entangling sublayer incorporating the hopping terms ($\imath$SWAP gates), followed by the other entangling sublayer containing the interaction terms (controlled-$R_z$ gates), and then a one-qubit sublayer composed of $R_z$ gates.

The hopping sublayer consists of $N(L - 1)$ $\imath$SWAP gates and the interaction sublayer is composed of $L(N - 1)$ controlled-$R_z$ gates. Each $\imath$SWAP gate introduces three CNOT gates \cite{Vatan_2004}, while each interaction term consisting of a controlled-$R_z$ gate decomposes into two CNOT gates \cite{2005quant.ph..4100M}. Therefore, the total number of decomposed CNOT gates is equal to $5NL - 3N - 2L$. Looking at the circuit depth, the hopping sublayer garners a depth of $3(L - 1)$, with the interaction sublayer achieving a depth of $2(N - 1)$. Hence, the effective depth of the parameterized quantum circuit is equal to $2N +3L - 5$. Now counting the number of parameters, the hopping sublayer consists of $N(L - 1)$ parameters while the interaction sublayer contains $L(N - 1)$ parameters --- since the controlled-$R_z$ gates acting from each color to the next can be done in parallel --- with each one-qubit sublayer having $NL$ parameters. Thus, each ansatz layer incorporates $3NL - N - L$ parameters, along with the initial layer of one-qubit $R_z$ gates acting after the $X$ gates (used to create a spin), introducing an extra $N_p$ parameters at the start of the circuit. As a specific example, referring to Fig. \ref{fig:circuit} using one layer, we obtain a total of 24 parameters.

In each instance of the VQE, we considered an initial state starting in the computational basis equating to the spin configuration of the intended model. In fact, this ansatz allows for the simple notion of error detection to be implemented: by checking the Hamming weight (the number of 1s in the binary result) of the output and confirming that it is conserved with respect to each individual spin color. If a non-conserved binary string is outputted, then one can be certain that an error occurred during computation and the result of the VQE can be disregarded.

\subsection{Measurement}

By decomposing the Hubbard Hamiltonian given in Eq. \eqref{eq:mapr} into Pauli-strings, we end up with the following:
\begin{align}
H = -\ &\sum\limits_{i, s, r}t_r\left(e^{\imath\frac{2\pi\phi}{L}} \sigma_{i+sL}^+\sigma_{i+r+sL}^- + \text{h.c.} \right)\prod_{j=i+1+sL}^{i+r-1+sL}\sigma_{j}^z \nonumber\\
+\ &\dfrac{U}{4}\sum\limits_{i, s<s'}\sigma_{i+sL}^z\sigma_{i+s'L}^z +\frac{1}{4}\sum\limits_{i, s, s', r} V_r\sigma_{i+sL}^z\sigma_{i+r+s'L}^z\nonumber\\
-\ &\frac{1}{4}\left( N\lambda_{R_V}(\Vec{V}) + (N - 1)U \right) \sum\limits_{i, s} \sigma_{i + sL}^z\nonumber\\
+\ &\dfrac{NL}{8}\left( N\lambda_{R_V}(\Vec{V})+ (N - 1)U \right) \,,
\label{eq:map2}
\end{align}
with
\begin{align}
    \lambda_{R_V}(\Vec{V}) = \sum_{r = 1}^{R_V} g_L(r) V_r \,, \\
    \Vec{V} = \left\{V_1, V_2, \dots, V_{R_V}\right\} \,,
\end{align}
and
\begin{equation}
    g_L(r) = 
    \begin{cases}
        2 \text{ if } r < \frac{L}{2} \,, \\
        1 \text{ if } r = \frac{L}{2} \,.
    \end{cases}
\end{equation}
Note that $\lambda_R = \sum_{r = 1}^{R} g_L(r)$ counts the number of edges in a circulant graph \cite{circulant} having edges of up to distance $R$, and at most $\left\lfloor \frac{L}{2} \right\rfloor$, representative of the layout of a 1D ring lattice.

Thus, the Hubbard Hamiltonian consists of $NL\lambda_{R_t}/2$ $\sigma^x\sigma^x$, $\sigma^y\sigma^y$, $\sigma^x\sigma^y$ and $\sigma^y\sigma^x$ terms each, $NL(N(1 + \lambda_{R_V}) - 1)/2$ $\sigma^z\sigma^z$ terms, $NL$ $\sigma^z$ terms, and one constant term, adding to a total of $NL(4\lambda_{R_t} + N(1 + \lambda_{R_V}) + 1)/2$ terms, which equates to $3NL(N + 3)/2$ for nearest-neighbor hopping and interaction models as in Eq. \eqref{eq:map1}. This means that naively taking separate energy measurements for each term in the Hamiltonian would prove to be severely error-prone. One solution to this problem is incorporating commuting sets of observables and measuring them in parallel \cite{Jena2019, Zhao2020}. Immediately, it can be observed that the $\sigma^z\sigma^z$ and $\sigma^z$ terms can be measured in parallel, since they are already diagonal operators (and thus commuting) in the computational basis. The simplest way is to take the original form of $(1 - \sigma^z_i)(1 - \sigma^z_j)/4 = \ket{11}\bra{11}_{ij}$ for the $\sigma^z\sigma^z$ terms and $(1 - \sigma^z_i)/2 = \ket{1}\bra{1}_{i}$ for the $\sigma^z$ terms, as given directly in Eq. \eqref{eq:map1}. Hopping terms can be divided into three sets: \textbf{a)} the even-odd hopping terms, i.e. the terms in Eq. \eqref{eq:hopping} acting on qubits 0-1, 2-3, 4-5, \dots \textbf{b)} the odd-even hopping terms, i.e. qubits 1-2, 3-4, 5-6, \dots, and lastly \textbf{c)} the closed hopping term acting on qubits 0-$(L-1)$, along with the $\sigma^z$ terms in between. To measure the hopping terms given in Eq. \eqref{eq:hopping}, a unitary operator that diagonalizes pairs of qubits in the hopping basis is given by
\begin{equation}
	U_L(\phi) = \left( \begin{array}{cccc}
	    1 & 0 & 0 & 0 \\
		0 & \frac{e^{-\imath\frac{\pi\phi}{L}}}{\sqrt{2}} & \frac{e^{\imath\frac{\pi\phi}{L}}}{\sqrt{2}} & 0 \\[1.8ex]
		0 & -\frac{e^{-\imath\frac{\pi\phi}{L}}}{\sqrt{2}} & \frac{e^{\imath\frac{\pi\phi}{L}}}{\sqrt{2}} & 0 \\[1.5ex]
		0 & 0 & 0 & 1
	\end{array} \right) \,.
	\label{eq:V}
\end{equation}
This unitary operator is responsible for transforming the hopping terms in Eq. \eqref{eq:hopping} to the diagonal basis $\ket{01}\bra{01} - \ket{10}\bra{10}$. It is significant to note that applying sets of this transformation before measuring effectively increases the circuit depth by a further three CNOT gates \cite{Vatan_2004}.

This implies that only four sets of measurements are needed to calculate the expectation value of the Hamiltonian containing only nearest-neighbor hopping. It is also significant to note that if the number of spins for each color is odd, and the number of sites is even, then one can ignore the $\sigma_z$ terms in between the closed hopping term, due to parity symmetry. This will enable the closed hopping term to fit in with the set of even-odd hopping terms, further reducing the total sets of measurements to three.

\subsection{Classical Optimization}

The BFGS method \cite{nocedal2006bfgs} is the classical optimization technique used in all of the statevector simulations. It was specifically chosen due to its relatively quick and accurate convergence, requiring a moderate amount of iterations to arrive at the ground-state energy, when compared with other optimization algorithms. For all the simulations, the threshold for convergence was defined to be a tolerance value of $10^{-5}$ for the gradient norm between one iteration and the next. However it is important to note that the BFGS algorithm would not be ideal in the case that the simulations were performed with sampling measurement outcomes rather than by taking exact energy measurements. For this reason, carrying out a simulation either on a classical computer or on a quantum computer, and sampling measurement outcomes, requires the use of a more sophisticated optimization technique, such as NFT \cite{NFT} (also known as Rotosolve \cite{Tilly2021}), to minimize the effect of the inherent statistical noise in the measurement results, as shown in \ref{S.sampling}.

In the case of sampling, the BFGS algorithm \cite{nocedal2006bfgs} requires sampling of both the gradient and the curvature of the parameter space, meaning that both the Jacobian and the Hessian need to be computed for each sample. The classical complexity also scales with the cube of the number of parameters of the variational ansatz. On the other hand, the NFT algorithm \cite{NFT} is a gradient-free optimizer, which only optimizes one parameter at a time. Due to analytical landscape of each parameter being a sinusoidal function, NFT uses this fact to perform sinusoidal fitting, which has constant time complexity. Consequently, at each iteration only three measurement are required for each parameter, achieving a linear scaling on the number of measurements.

\subsection{Complexity Analysis}

In this section we compile the complexity scalings of both the variational ansatz, as well as of the optimizers used in the simulations. Table \ref{tab:cs} highlights the computational scaling of the variational ansatz, while Table \ref{tab:ca} outlines the complexity analysis of VQE procedure using the different optimizers mentioned in this paper.

\begin{table}[h]
\centering
\begin{tabular}{||c c c||} 
    \hline
    CNOT Gates & Circuit Depth & Parameters \\
    \hline
    \hline
    $5NL - 3N - 2L$ & $2N + 3L - 5$ & $3NL - N - L$ \\ 
    \hline
    $\mathcal{O}(NL)$ & $\mathcal{O}(N + L)$ & $\mathcal{O}(NL)$ \\ 
    \hline
\end{tabular}
\caption{Computational scaling of the variational ansatz, where we highlight the number of decomposed CNOT gates, associated circuit depth, and the number of parameters in the ansatz.}
\label{tab:cs}
\end{table}

\begin{table}[h]
\centering
\begin{tabular}{||c c c c||}
    \hline
    Optimizer & Type & Measurement Complexity & Classical Complexity \\
    \hline
    \hline
    BFGS & Second-order & $S(g_1 + g_2) $ & $\mathcal{O}(p^3) = \mathcal{O}(N^3L^3)$ \\ 
    \hline
    NFT & Gradient-free & $\mathcal{O}(p) = \mathcal{O}(NL)$ & $\mathcal{O}(1)$ \\ 
    \hline
\end{tabular}
\caption{Complexity analysis of the different optimizers \cite{Tilly2021}, used in this paper, where we highlight the type of optimizer, and its corresponding measurement complexity, denoting the amount of measurements needed per iteration, as well as its classical complexity, representing the complexity of the optimizer for each iteration. $S$ denotes the number of samples taken per iteration, $g_1$ and $g_2$ represent the cost of evaluating the first-order and second-order gradients, respectively, and $p$ denotes the number of parameters in the circuit.}
\label{tab:ca}
\end{table}

\section*{References}

\bibliographystyle{unsrt}
\bibliography{ref}

\end{document}